\documentclass[aps,twocolumn,noshowpacs,superscriptaddress,floatfix,nofootinbib]{revtex4-1}
\usepackage[table]{xcolor}
\usepackage{quantikz}
\usepackage{enumerate}
\usepackage{amsmath,amssymb}
\usepackage{mathrsfs}
\usepackage[justification=justified]{caption}
\usepackage{xcolor}
\usepackage{verbatim}
\usepackage[normalem]{ulem}
\usepackage[rightcaption]{sidecap}
\usepackage{graphicx}
\usepackage{caption}
\usepackage{multirow}
\usepackage{amsthm,epsfig,tikz}
\usepackage{float}
\usepackage{multirow}
\usepackage[colorlinks,citecolor=black,urlcolor=black,linkcolor =black,bookmarks=false,hypertexnames=true]{hyperref} 
\usepackage{relsize}
\usepackage{svg}
\usepackage{float}
\makeatletter
\let\newfloat\newfloat@ltx
\makeatother
\usepackage{algorithm}
\usepackage{algcompatible}
\usepackage{subcaption}

\newcommand{\norm}[1]{\left\lVert#1\right\rVert}

\renewcommand{\part}[2]{\frac{\partial #1}{\partial #2}}

\usepackage{float}
\makeatletter
\let\newfloat\newfloat@ltx
\makeatother

\begin{document}

\title{Hybrid Quantum Graph Neural Network for Molecular Property Prediction}
\author{Michael Vitz}
\affiliation{Deloitte Consulting, LLP}

\author{Hamed Mohammadbagherpoor}
\affiliation{IBM Quantum}

\author{Samarth Sandeep}
\affiliation{Deloitte Consulting, LLP}

\author{Andrew Vlasic}
\affiliation{Deloitte Consulting, LLP}

\author{Richard Padbury}
\affiliation{IBM Quantum}

\author{Anh Pham}
\email{anhdpham@deloitte.com}
\affiliation{Deloitte Consulting, LLP}

\date{\today}

\begin{abstract}
To accelerate the process of materials design, materials science has increasingly used data driven techniques to extract information from collected data. Specially, machine learning (ML) algorithms, which span the ML discipline, have demonstrated ability to predict various properties of materials with the level of accuracy similar to explicit calculation of quantum mechanical theories, but with significantly reduced run time and computational resources. Within ML, graph neural networks have emerged as an important algorithm within the field of machine learning, since they are capable of predicting accurately a wide range of important physical, chemical and electronic properties due to their higher learning ability based on the graph representation of material and molecular descriptors through the aggregation of information embedded within the graph. In parallel with the development of state of the art classical machine learning applications, the fusion of quantum computing and machine learning have created a new paradigm where classical machine learning model can be augmented with quantum layers which are able to encode high dimensional data more efficiently. Leveraging the structure of existing algorithms, we developed a unique and novel gradient free hybrid quantum classical convoluted graph neural network (HyQCGNN) to predict formation energies of perovskite materials. The performance of our hybrid statistical model is competitive with the results obtained purely from a classical convoluted graph neural network, and other classical machine learning algorithms, such as XGBoost. Consequently, our study suggests a new pathway to explore how quantum feature encoding and parametric quantum circuits can yield drastic improvements of complex ML algorithm like graph neural network.
\end{abstract}

\maketitle

\section{Introduction}
With a focus on predicting properties of complex materials, this manuscripts derives a novel and unique gradient-free hybrid quantum graph neural network algorithm. Through the application of predicting the formation energies of perovskites, the hybrid quantum-classical convoluted graph neural network (CGNN) regression model yielded an R-squared value of $R^2 = 0.674$, displaying an algorithm competitive with a pure classical CGNN. In particular, our work demonstrated the utility of quantum algorithms to augment a complex ML class of algorithms, CGNN, in performing prediction on the complex dataset of perovksites \cite{Perovskite_dataset}. Of particular interest is the utilization of gradient-free techniques to train the statistical model. The implementation of gradient-free optimization sub-processes is beyond the scope of a previous study on quantum graph neural network for materials research \cite{{ma16124300}}, as well as other studies that explore different quantum graph neural network (QGNN) algorithms \cite{ai2023decompositional, verdon2019quantum, QGNN_particle_tracking}.

The remainder of the manuscript is organized as following. Section \ref{subsec:intro-ml} gives an overview of how classical machine learning algorithms used within the field of materials science, with a special focus on classical graph neural network, while \ref{subsec:intro-qml} provides a brief review of previous studies of how quantum machine learning algorithms can be used to accelerate material discovery, and \ref{subsec:intro-motivation} presents the motivation for the research. Section \ref{sec:method} provides a detailed description of our feature engineering process by extracting the relevant descriptors of perovskites, the architecture of our hybrid quantum-classical graph neural network, non-gradient based optimization, the XGBoost algorithm, and the training and evaluation of our results. Section \ref{sec:rslts&dis} summarize and discuss the results obtained from our classical, XGBoost and hybrid algorithm, and section \ref{sec:conclusion} provides the concluding remarks of our study.

\subsection{Overview of Machine Learning Approach for Material Discovery}\label{subsec:intro-ml} In-silico simulations have accelerated the discovery of new materials and molecular candidates by performing atomistic simulation to predict their properties. However, an open challenge within the field of material simulation is the polynomial increase in computational time as the molecules or solid state materials become larger in sizes \cite{scaling_QC_classical, Ab_initio_QC}. Furthermore, due to the limitation of certain quantum mechanical approximation of the exact interaction within the materials' atomic structure \cite{DFT_limit}, certain materials properties are difficult to calculate accurately. These characteristics result in the reliance to extract these properties using laboratory techniques, and such laboratory measurement processes can be prohibitively expensive. As a consequence, data-driven techniques have emerged as an important toolkit to accelerate the material design process, in addition to quantum simulation and experimental studies. 

Moreover, the advancement of data-driven strategies in material design can also be attributed to the increasingly availability of data-sets \cite{MP_project} based on high throughput screening studies as well as experimental database \cite{Experimental_database}. This has enabled the materials science community to apply complex machine learning techniques to data mine and potentially extrapolate the complex inter-dependencies of structure and property in materials to make predictions of new materials with novel properties. 

In fact, through various applications, machine learning models have demonstrated their utilities in correlating the complex interplay between structure-properties relationship, which enables these models to  predict materials' properties such as atomization energies \cite{descriptorPRL2015}, formation energies \cite{Bartel2020}, bulk modulii \cite{Isayev2017}, and band gap energies \cite{Zhuo2018}  at low computational cost. In addition, the emergence of deep learning also allows for the prediction of quantum mechanical properties of materials such as the electronic ground states of many body problems \cite{PhysRevResearch.2.033429, DeepQMC}, and density of states \cite{Chen2023} with a high level of accuracy. Interestingly, in comparison with explicit calculation using state-of-the art quantum chemistry methods, this accuracy was computed with a significantly reduced run-time.

In ML, feature engineering is an essential sub-process in the workflows. For materials science, the process of engineering features involves the extraction of various chemical descriptors, which then are columns in a data frame format and ingested into various machine learning models. Among many classical machine learning models, graph neural networks (GNNs) represent the best-in-class algorithm for materials' property prediction due to their high learning capabilities \cite{GNN_review} from the latent representation derived the materials' chemical descriptors. GNNs utilize the graph structure to encode data features, on top of a neural network architecture which extracts the most relevant features of the nodes and edges to perform a predictive task. 

For the materials science discipline, graph representation of atomic and bonding features represent a more natural expression of materials since the atomic features can now be encoded in the nodes, while bonding properties can be encoded as edge features, and the connectives between the atoms can be expressed as the number of edges within a graph. Such a representation was first used by Xie et al. \cite{xie2018crystal} to encode a unit cell of solid in the form of a crystal graph convolution neural network (CGCNN) to predict properties of solids. Subsequently,  other more advanced GNN models have been developed, like MegNet \cite{Chen2019} which includes more global state inputs like pressure, temperature and entropy, and GATGN \cite{GATGN} which includes local and global attention layer in the neural network to increase the expressivity of the model.

While machine learning algorithms have demonstrated utilities in predicting materials properties to various degree of accuracy, there are potential pitfalls for machine learning in the field of materials science. In particular, the lack of training data for certain exotic compounds. In addition, the predictive power of a machine learning model can be strongly dependent on the feature extraction from the data of the material. Such a process requires expert knowledge and intuition \cite{Wagner_feature_knowledge}. To compensate for the a priori knowledge, various automated algorithms \cite{Automate_feature} have been proposed to extract a vast array of features, but are unable to elongate the model training time due to increasing requirement of available data points with increasing number of features \cite{Sample_vs_feature}. Another potential drawback is that many GNNs utilize a message passing scheme to learn the relevant features iteratively. Such an approach can suffer the non-convergence problem \cite{liu2021eignn}.

\subsection{Quantum Machine Learning Application in Materials Science} \label{subsec:intro-qml} 
To compensate for the the potential shortcomings mentioned in the previous subsection, an alternative solution is to utilize quantum computing in order to perform machine learning training and prediction in an exponentially large Hilbert space. By taking advantage of the quantum mechanical properties, quantum machine learning (QML) algorithms were mathematically shown to require less training data to build a predictive model in comparison to their classical counterpart \cite{Caro2022}. In addition, QML also allows for a better mapping high-dimensional feature space of classical data, which enables better prediction for certain data structures \cite{Huang2021}. Recently, it has been suggested that certain QML model can also emulate the long-range correlation in classical data through quantum contextuality \cite{PRXQuantum.4.020338}, thus allowing the models to be both explainable and accurate. 

Even within the NISQ area, various QML algorithms have displayed promising results in different application for materials science and molecular engineering. For instance, quantum support vector machines (QSVM) has been applied in different areas of the drug discovery process such as classifying drug toxicities \cite{Bhatia2023}, as well as identifying binding sites in virtual screening \cite{Mensa_2023}. The more sophisticated learning task of quantum neural network (QNN) through the applications of generating new small molecules \cite{Kao2023, Li_QGAN_molecule}, force fields parameter \cite{2023arXiv231111362L} for molecular dynamic simulation, and the electronic ground states \cite{Wiebe_QGAN}, demonstrated an advantage using quantum computing. In the context of graph machine learning (GML), the augmentation of GML with feature encoding in the form of a Hamiltonian representation can outperform classical GML with less features \cite{PhysRevA.107.042615}.

\subsection{Our Motivation}\label{subsec:intro-motivation} 
To leverage the potential advantage of quantum properties, various hybrid models combining quantum and classical neural networks have been proposed, and many of them have been applied successfully to perform classification and prediction of materials and molecules' properties \cite{Reddy_2021, battery_prediction, Domingo2023, ma16124300}.  Within this context, we explored the potential enhancing statistical model performance for materials science prediction by augmenting traditional graph convoluted neural networks with quantum feature encoding and with quantum neural network layer. Specifically, we utilized an amplitude encoding method to encode the adjacency matrix in our quantum circuit in conjunction with a trainable ansatz. In addition, to avoid the vanishing gradient problem \cite{vanishing_gradient, McClean2018} in variational algorithms, we incorporated a gradient-free optimization method \cite{Trajanov_2022, nevergrad} in our hybrid algorithm. Our results show that the hybrid model is competitive against classical algorithms but does not improve on the results. Particularly, using simplified features composed mainly of electronic properties, we compared results obtained from a pure classical convoluted graph neural network \cite{li2020deepergcn}, and state-of-the-art gradient boosting algorithms, XGBoost \cite{chen2016xgboost}, for predicting properties like formation energies of perovskites materials. Consequently, the competitive predictive power of the hybrid model through including a quantum encoding of the materials' descriptors and graph connectivities, and the utilization of non-gradient based optimization suggest that there is potential improvement of performance in QGNN models through the development of new feature mappings of graphs .

\section{Methodologies}\label{sec:method}
\subsection{Chemical Descriptors and Chemical Bonding as Features}
\subsubsection{Training data and feature extraction}
\begin{figure}[htbp!]
\includegraphics[width=8cm]{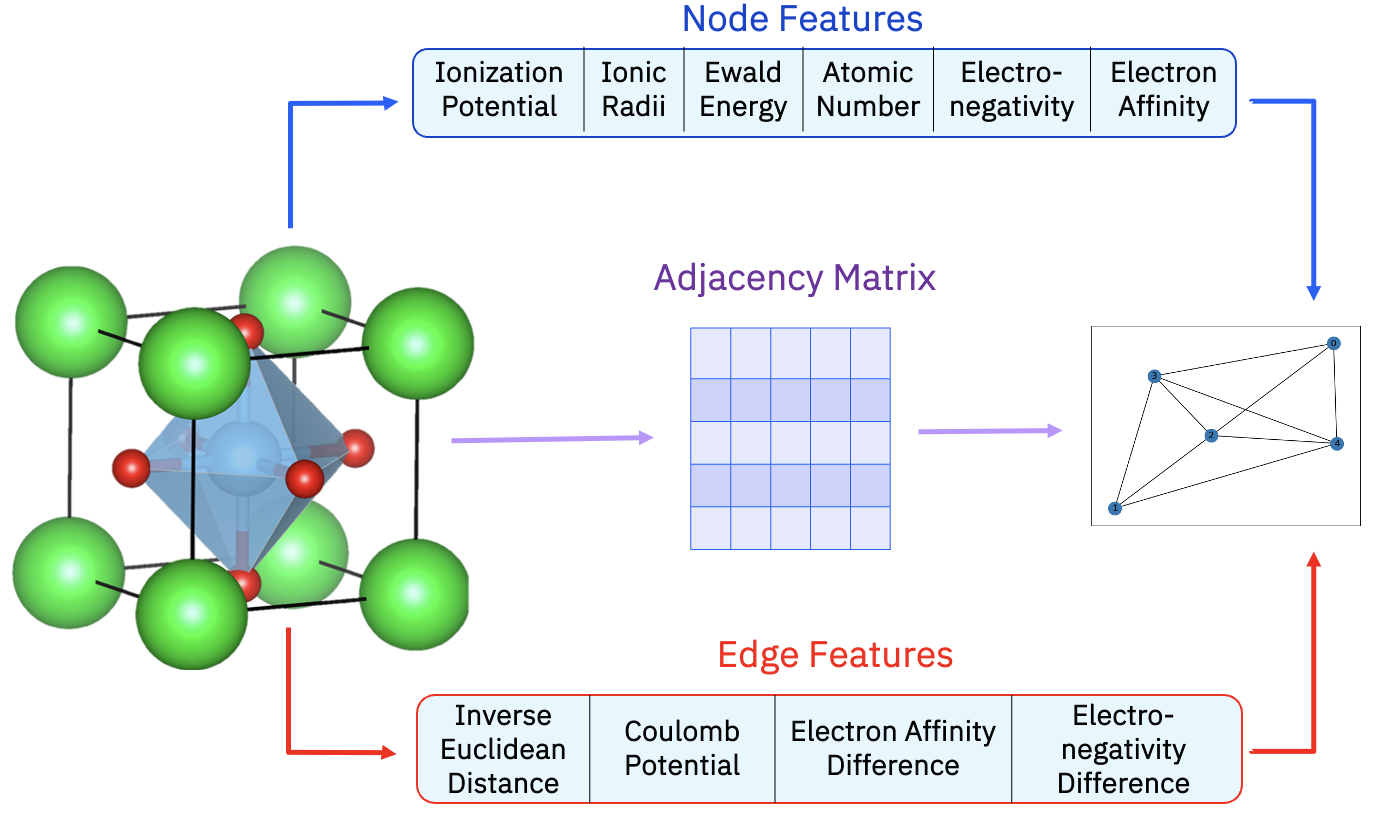}
\caption{Featurization overview}
\label{featurizer-overview}
\end{figure}
The perovskite dataset \cite{Perovskite_dataset} was downloaded from the MatBench database \cite{Dunn2020} and cleaned and featurized. To clean the data, we restricted attention to oxide perovskite which have a formula of $ABO_3$. In our study, we aimed to predict the formation energy of our oxide perovskite using various machine learning models, and compared the results with density functional theory (DFT) calculations. To accomplish this task, PyMatGen \cite{Pymatgen_Ong} was used to perform the feature extraction from the materials data-set. We extracted the electronic properties of the individual atoms as our node features. Specifically, these properties are: atomic number, Ewald energy, electronegativity, electron affinity, ionization potential, cationic and anionic radii. Chemical descriptors like electronegativity and electron affinity have been demonstrated to be important in machine learning models used to predict the formation energy of binary compounds \cite{mao2021prediction}. Other descriptors like ionization potential and cationic and anionic radii are also important descriptors in predicting phase stability in periodic system \cite{descriptorPRL2015}. While Ewald energy for the individual site in the perovskite structures can capture the long range interaction \cite{EwaldML} which can be important in a graph neural network model \cite{Ewaldmoleculargraph2023}. \\
\indent To encode the features for the edges, we extract the properties that connect the individual atoms such as the inverse of the Euclidean distances, the coulomb potential between the neighbouring atoms, the electronegativity and electron affinity differences between the two atoms. For the Coulomb matrix descriptor, we only extracted the values of the off-diagonal element which represent the interaction between two different nuclei. 

\subsubsection{Encoding of chemical bonding into graph structures}\label{subsec:method-hybrid}
As shown in figure \ref{featurizer-overview} each of the perovskite crystals from the Castelli perovskites dataset was translated into a graph by mapping the chemical features of each atom onto the node features. Connectivities between nodes were mapped as bond relationships between the atoms based on the adjacency matrices found for each atom and how they are connected to each other. By incorporating the spatial geometric structure of the materials into  graph in the form of adjacency matrices have been suggested to enhance the predictive power of various classical graph neural network models \cite{Cheng2021}. 

\subsection{Hybrid Quantum-Classical Convoluted Graph Neural Network (HyQCGNN)}
\subsubsection{Classical layer: Generalized Graph Convolution Neural Network (GENConv)}

For a classical graph neural network layer, we use the GENConv model \cite{li2020deepergcn}, which has a number of desirable properties. Two techniques in particular increase the expressive power of a GNN substantially. 
\begin{enumerate}
    \item GENConv uses a generalized aggregation function, with trainable parameters allowing the model to learn the optimal aggregation method. For example, we used SoftMax with a learnable “temperature” parameter. For low values of inverse temperature, this behaves like mean aggregation and at high inverse temperatures it behaves like max aggregation. This technique is well-established in numerous fields of machine learning \cite{HE201880, piao2022sparse}. 
    \item GENConv uses a message normalization technique to normalize features of aggregated messages that combines them with other features (like edge features) during the update phase. This is accomplished through a MLP\footnote{Note that in our case, we use a single fully connected layer, so our architecture is not truly an MLP, although GENConv allows for multiple layers.} and a learnable scaling factor $s$, yielding the vertex update function:
    \begin{eqnarray*}
        \mathbf{h}_{\nu}^{(l+1)} & = & \phi^{(l)}(\mathbf{h}_{\nu}^{(l)},\mathbf{m}_{\nu}^{(l)}) \\
        & = & \mathbf{MLP}\left(\mathbf{h}_{\nu}^{(l)}+s\cdot \norm{\mathbf{h}_{\nu}^{(l)}}_2 \cdot \frac{\mathbf{m}_{\nu}^{(l)}}{\norm{\mathbf{m}_{\nu}^{(l)}}_2}\right)
    \end{eqnarray*}
\end{enumerate}

It is worth mentioning that not all GNNs permit the use of edge features at all, while the MLP implemented by GENConv includes them in a particularly powerful way.

Classical GNNs (whether GENConv or other) can make use of the backpropagation techniques \cite{Rumelhart1986} established across the field of neural network architectures to calculate gradients. In quantum machine learning, parameter-shift rule \cite{Banchi2021measuringanalytic} has been shown to be a solution to compute the gradients in variational quantum circuits to optimize trainable parameters. However, this method has been shown to be less efficient than the classical backpropagation in traditional ML \cite{PRXQuantum.3.030101}. As a result, in our study we explored the use of gradient-free approach for both classical and hybrid model as detailed in  \ref{gradient-free-opt}, to facilitate like-to-like comparisons.

\subsubsection{Quantum Layer}\label{quantum-layer}

The encoding of a feature vector $\Vec{x}=(x_0,x_1,\ldots,x_N)$ into a quantum state requires initializing the state to have amplitudes equal to each feature vector element, as below:
\begin{equation}
    \label{amplitude-encoding}
|\psi\rangle  =  \frac{1}{\sqrt{N}}\sum_{i=0}^N |x_i\rangle
\end{equation}

\begin{figure}[!ht]
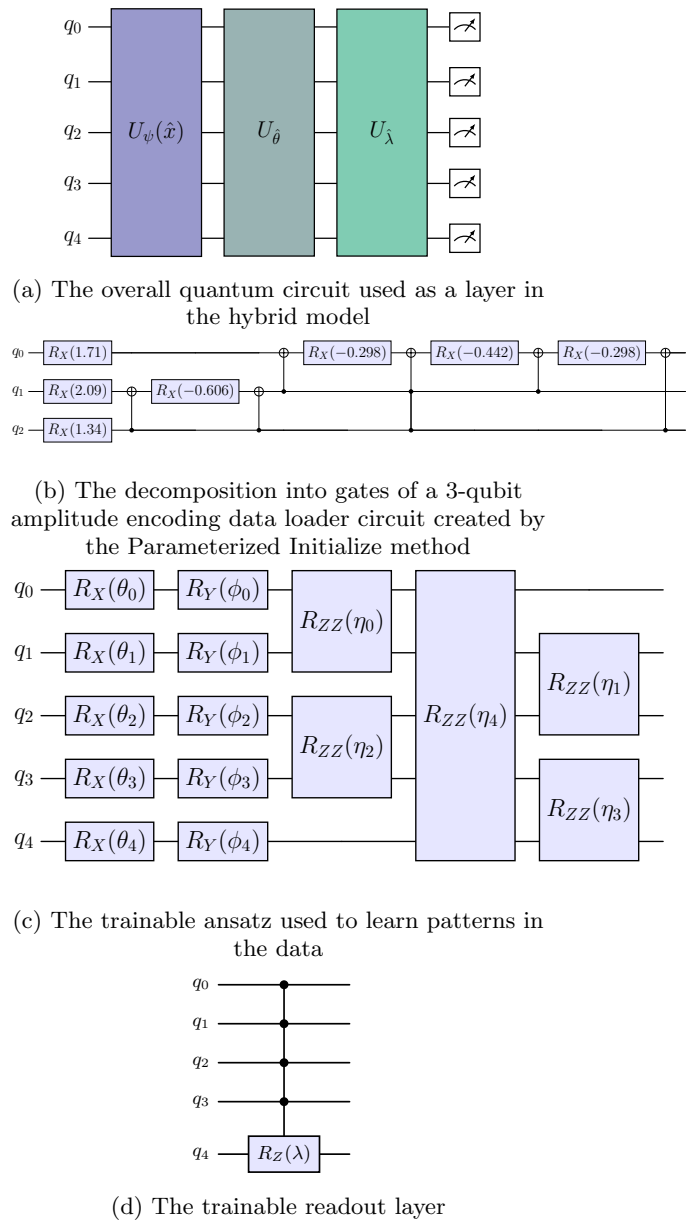

    \centering
    \begin{subfigure}[a]{0.4\textwidth}
    \centering
     \scalebox{0.6}{\input{high-level-circuit}}  
 
    \caption{The overall quantum circuit used as a layer in the  hybrid model}\label{full-quantum-circuit}
    \end{subfigure}
    \hfill
    \begin{subfigure}[b]{0.4\textwidth}
    \centering
     \scalebox{0.4}{\input{amplitude-encoding}}  

  \caption{The decomposition into gates of a 3-qubit amplitude encoding data loader circuit created by the Parameterized Initialize method} \label{decomposed-amplitude-encoding}
    \end{subfigure}
    \hfill
    \begin{subfigure}[c]{0.4\textwidth}
    \centering
    \tikzset{every picture/.style={scale=0.3}}
     \scalebox{0.65}{\input{variational-ansatz}}  

    \caption{The trainable ansatz used to learn patterns in the data} \label{trainable-ansatz}
    \end{subfigure}
    \begin{subfigure}[d]{0.4\textwidth}
    \centering
    \tikzset{every picture/.style={scale=0.3}}
     \scalebox{0.8}{\input{variational-readout}}  

    \caption{The trainable readout layer} \label{trainable-readout}
    \end{subfigure}
\caption{Quantum circuit implemented in our hybrid QGCNN model which includes a data loading layer using amplitude encoding, a variational layer, and a trainable readout layer.  For space considerations the amplitude encoding layer is restricted to 3-qubits
}
\label{fig:circuit-diagrams}
\end{figure}

The circuit form of the amplitude encoding is shown in figure \ref{decomposed-amplitude-encoding}, for the case where $N_{qubits}=3$ for illustrative purpose. The quantum circuit was implemented using the RawFeatureVector function \cite{Shende_2006} within Qiskit. In order to model the perovskite structure of our study, we utilized 5 qubits since the chemical structures contain 5 atoms within a primitive unit cell. Thus, they can be mapped to 5 nodes within our adjacency matrix.

By unrolling our adjacency matrix into a 1-D vector, we can use the equation \ref{amplitude-encoding} to determine state amplitudes. After loading the data in this way, we applied a trainable ansatz layer, which allows us to adjust incrementally the parameters. The general form of the ansatz is shown in figure \ref{fig:circuit-diagrams}

We then applied a readout layer, which also includes trainable parameters, rotating the qubits around the $X,Y,Z$ axes. Since the trainable rotations occur right before measurement, the readout layer can be thought of as learning which observable to measure for optimal information extraction from the circuit. 

\subsubsection{Non-Gradient-Based Optimization}\label{gradient-free-opt}
Guided by recent research of exploring gradient-free optimization algorithm for quantum neural network model \cite{kulshrestha2023learning, wiedmann2023empirical}, we performed our optimization using non-gradient-based approaches.

We used an open-source gradient-free package from Facebook research, called Nevergrad \cite{Trajanov_2022,nevergrad}. This package includes numerous optimization algorithms. Their default algorithm, called NGOpt is a carefully tuned black-box optimization, sometimes called an ``algorithm selection wizard''. Such selection wizards auto-select optimization procedures based on details of the search space, defined at the outset of optimization. Extensive research has gone into these selection wizards, and their performance is difficult to beat. Furthermore, Nevergrad has been shown to be an effective algorithm to optimize the cost function for variational method like the quantum approximate optimization algorithm \cite{baker2022wasserstein, Yao2020PolicyGB}.

\subsubsection{Algorithm Workflow}

The steps for optimization of our hybrid model are as follows.
\begin{enumerate}
    \item The optimizer is queried to produce a set of values for the model parameters, which will include subsets for both the classical and quantum layers. 
    \item The classical model parameters are loaded into the GENConv layer.
    \item The GENConv model then transforms the input graph producing an intermediate graph. This intermediate graph has one node feature and one edge feature, distilled by GENConv. In other words, GENConv is performing trainable dimensionality reduction on the node and edge features. 
    \item This intermediate graph, which is represented by a (symmetric) weight matrix, is unrolled into a 1D vector, denoted $\vec{x}$.
    \item The vector $\vec{x}$ is loaded into a quantum circuit via amplitude encoding along with the quantum subset of trainable parameters.
    \item A small set of observables are measured for the circuit, then scaled and compared to the target value of our data-set.
    \item The difference between the target and the model prediction is reported back to the optimizer, which updates accordingly.
    \item An updated set of parameters is requested from the optimizer.
\end{enumerate}

\subsection{XGBoost}\label{subsec:method-xgb}
The XGBoost algorithm \cite{chen2016xgboost}, using gradient boosting on decision trees, is by now a well-established and well-understood model. We investigated the performance of XGBoost for the regression task of predicting formation energy from crystal graphs, both to serve as a baseline and to guide the featurization of the crystal graphs. In order to accomplish this, we used the fact that our graphs have homogeneous size and connectivity to flatten the graphs into a simple tabular data-set. This allowed us also to select features based on the feature importances reported by the XGBoost model after training.

\subsection{Training, Evaluation of Results, and Software Implementation}\label{subsec:method-eval}

Our neural networks were defined using pytorch \cite{paszke2019pytorch}, pytorch-geometric \cite{fey2019fast}. When defining circuits, we used Qiskit and the connectors that Qiskit exposes to pytorch neural network modules. Our hybrid quantum-classical model, pure GENConv, and XGBoost were trained with 196 data points, and the best model was selected by evaluating performance on validation set with 25 samples. Final evaluation was done on a test set containing 25 samples, which was not used in either model training or selection.

The results of our HyQCGNN model were obtained using the Qiskit Runtime environment and qiskit estimator primitive \cite{johnson2023ibm}. Specially, we utilized the the Estimator Quantum Neural Network (EstimatorQNN). EstimatorQNN is a hybrid neural network architecture that combines classical and quantum elements. In this structure, the quantum component, referred to as the feature map, transforms classical data into quantum states. The EstimatorQNN in general, is a neural network designed to process a parameterized quantum circuit with assigned parameters for input data and/or weights, along with an optional observable(s), generating the corresponding expectation value(s) as its output. EstimatorQNN from Qiskit leverages the Estimator primitive and allows users to combine parameterized quantum circuits with quantum mechanical observables, and can be connected to pytorch through TorchConnector. TorchConnector takes a neural network and makes it available as a PyTorch Module. The resulting module can be seamlessly incorporated into PyTorch classical architectures and trained jointly without additional considerations, enabling the development and testing of novel hybrid quantum-classical machine learning architectures. While applying TorchConnector, the user can get access to all of the well defined optimizer algorithms and pre-defined loss functions. 

All the models were trained using our internal cluster with 64 CPU cores and simulated on Qiskit QASM simulator. Feature importance analysis was also performed using XGBoost to understand how the different materials descriptors included in our data-set affect the predicted property of formation energies. Such analysis is done to further verify that we extracted the relevant features for our machine learning models, and their impacts are physically relevant. 

\section{Results and Discussion}\label{sec:rslts&dis}

Both the classical and the hybrid models were trained for 2000 iterations, using the NGOpt algorithm of Nevergrad. After every iteration, the model was evaluated on a validation data-set, and the best-performing model was recorded for later use. At the end of the 2000 iterations, the model that performed best on the validation set was evaluated on a test set, on which the model was not trained, and which played no part in model selection. We then plot the true formation energy values against the predicted values and perform a simple linear fit. The $R^2$ values for this fit are the figures of merit for the corresponding model.

\begin{figure}[!ht]
    \centering
    \begin{subfigure}[a]{0.4\textwidth}
    \centering
    \includegraphics[width=1.0\textwidth]{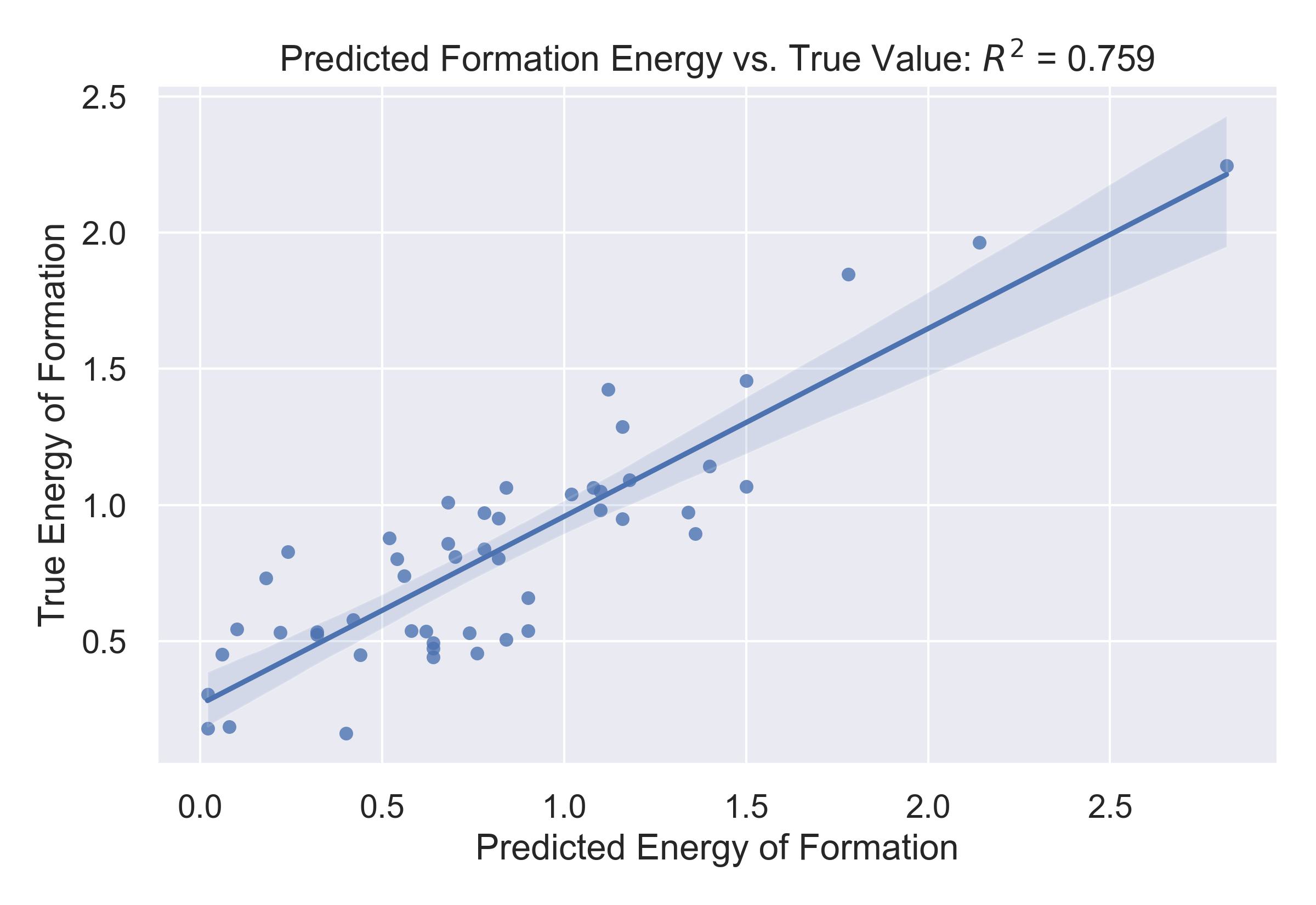}\caption{GENConv model performance}\label{classical-gnn-results}
    \end{subfigure}
    \hfill
    \begin{subfigure}[b]{0.4\textwidth}
    \centering
    \includegraphics[width=1.0\textwidth]{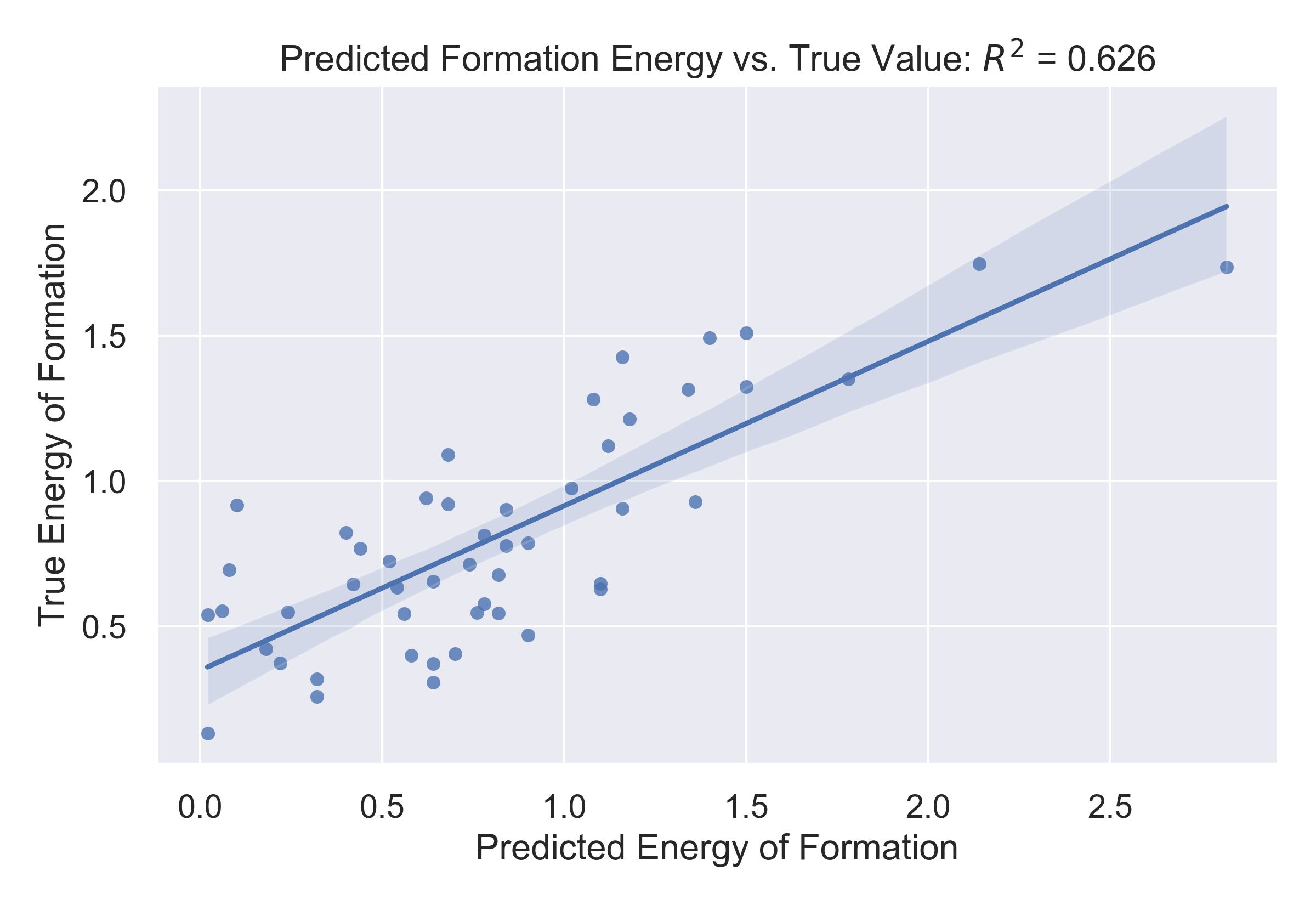}\caption{Hybrid model performance} \label{hybrid-gnn-results}
    \end{subfigure}
    \hfill
    \begin{subfigure}[c]{0.4\textwidth}
    \centering
    \includegraphics[width=1.0\textwidth]{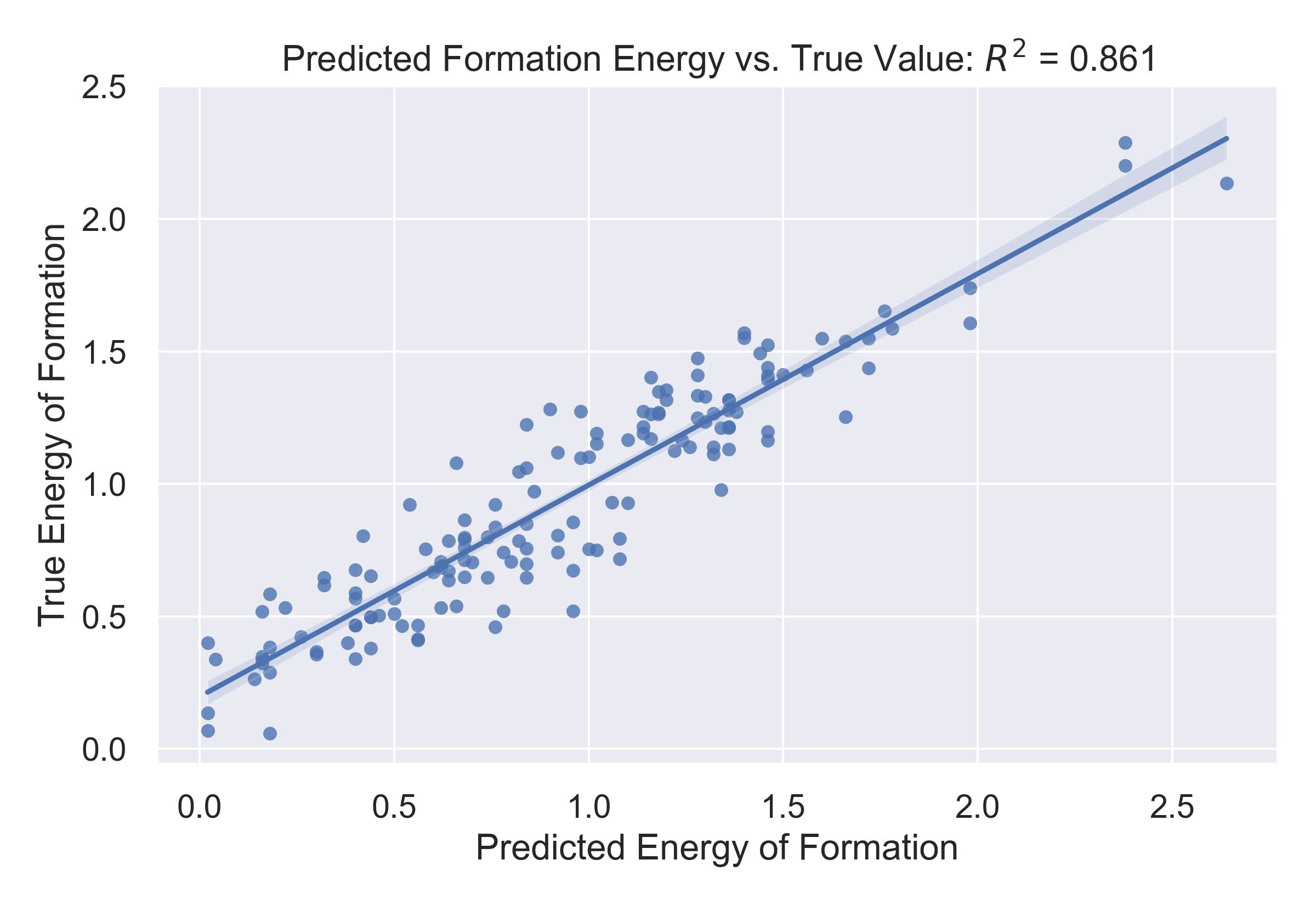}\caption{XGBoost model performance}\label{xgboost-r2-score}
    \end{subfigure}
\caption{Plot and associated $R^2$ value for the true formation energy vs different models' prediction of formation energy.}
\label{fig:model-r2-values}
\end{figure}

The results for both classical and hybrid methods are illustrated in figures \ref{classical-gnn-results} and \ref{hybrid-gnn-results}

\begin{figure}[htbp!]
\includegraphics[width=8cm]{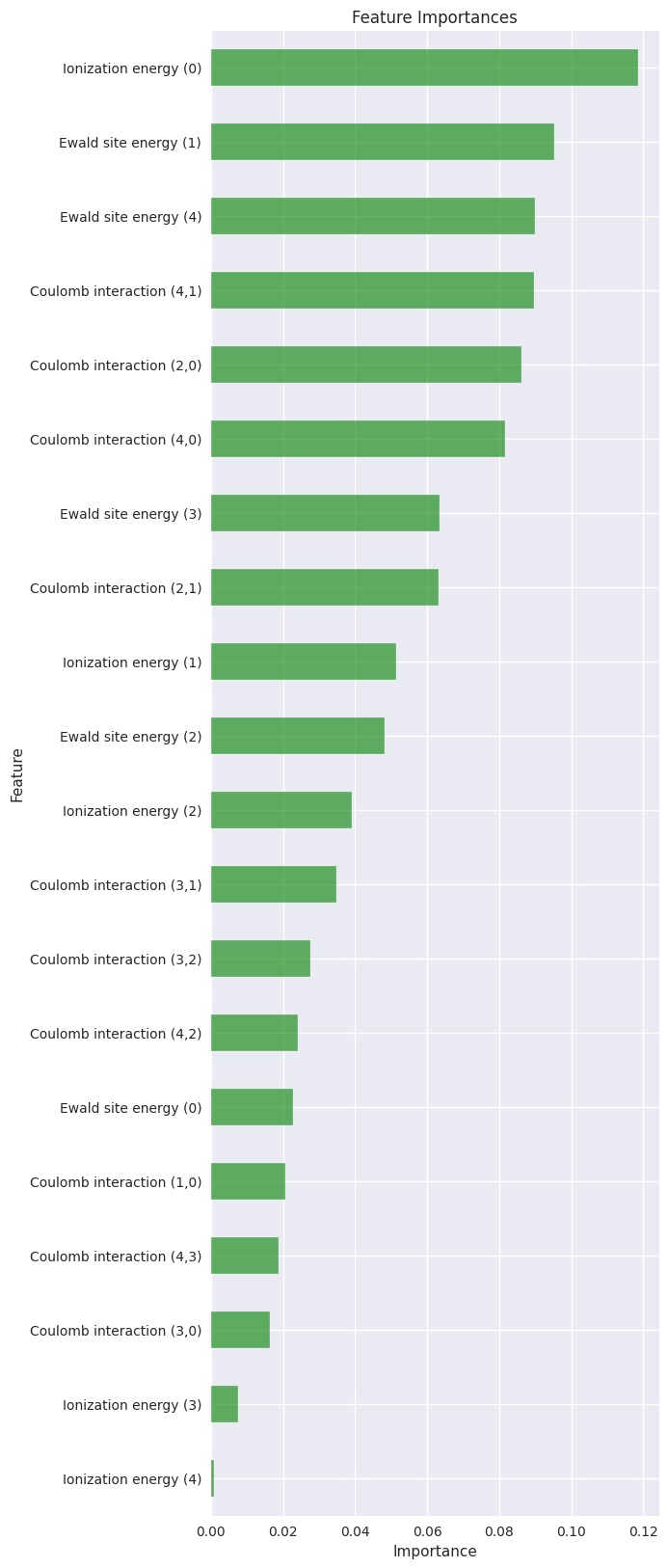}
\caption{Feature importances reported by the XGBoost algorithm. The parenthesis indicate which atom or pair of atoms the feature applies to.}
\label{feature-importances}
\end{figure}

The $R^2$ score obtained for the classical GNN is comparable to the XGBoost model. This illustrates that, despite the handicap of not using gradients in the optimization, the GNN architecture is able to perform quite well. In addition to the performance metric of $R^2$, XGBoost was also used to perform the feature importance analysis as shown in Fig. \ref{feature-importances}. The results suggest that the most relevant feature affecting the formation energy is the first ionization energy of site A in a perovskite material. The importance of ionization energies have been known to be an important descriptor to predict the stability of oxide perovskites \cite{Wexler2021, C1EE02377B}, and halide peroskites \cite{Zheng2017}. 

On the other hand, the $R^2$ score predicted by the hybrid quantum-classical model is slightly lower than the classical GNN and XGBoost results. This illustrates that the advanced methods used in the GENConv are more powerful than the quantum methods employed here. This is to be expected, as GENConv is among the most powerful GNN models available. Nevertheless, a competitive $R^2$  suggests that a hybrid quantum GNN is a viable model to perform prediction of complex materials' properties. A potential improvement is to perform further research in area of graph embedding within the Hilbert space to design a full quantum graph convoluted neural network as suggested by Hu et al. \cite{QuGCN}. However, such an investigation is not within the scope of this study and will be explored in subsequent research. 

\section{Conclusion}\label{sec:conclusion}

In this work, we have presented an implementation of a hybrid graph convolutional neural network that incorporates a quantum layer. This quantum layer includes both amplitude encoding and a trainable ansatz. Furthermore, we showed the feasibility of using non-gradient-based optimization for training this hybrid model. We report the figures of merit and compare to baseline scores obtained from advanced classical techniques. Although the classical techniques retain an advantage at present, the increasing maturity of quantum hardware, alongside the native $2^N$ scaling enabled by quantum amplitude encoding, suggests that this is a potentially fruitful avenue for continued research.

\subsubsection*{Disclaimer (Deloitte)}
About Deloitte: Deloitte refers to one or more of Deloitte Touche Tohmatsu Limited, a UK private company limited by guarantee (“DTTL”), its network of member firms, and their related entities. DTTL and each of its member firms are legally separate and independent entities. DTTL (also referred to as “Deloitte Global”) does not provide services to clients. In the United States, Deloitte refers to one or more of the US member firms of DTTL, their related entities that operate using the “Deloitte” name in the United States and their respective affiliates. Certain services may not be available to attest clients under the rules and regulations of public accounting. Please see  www.deloitte.com/about to learn more about our global network of member firms.

Deloitte provides industry-leading audit, consulting, tax and advisory services to many of the world’s most admired brands, including nearly 90\% of the Fortune 500® and more than 8,500 U.S.-based private companies. At Deloitte, we strive to live our purpose of making an impact that matters by creating trust and confidence in a more equitable society. We leverage our unique blend of business acumen, command of technology, and strategic technology alliances to advise our clients across industries as they  build their future. Deloitte is proud to be part of the largest global professional services network serving our clients in the markets that are most important to them. Bringing more than 175 years of service, our network of member firms spans more than 150 countries and territories. Learn how Deloitte’s approximately 457,000 people worldwide connect for impact at  www.deloitte.com.

This publication contains general information only and Deloitte is not, by means of this [publication or presentation], rendering accounting, business, financial, investment, legal, tax, or other professional advice or services. This [publication or presentation] is not a substitute for such professional advice or services, nor should it be used as a basis for any decision or action that may affect your business. Before making any decision or taking any action that may affect your business, you should consult a qualified professional advisor.
Deloitte shall not be responsible for any loss sustained by any person who relies on this publication. Copyright © 2023 Deloitte Development LLC. All rights reserved.

\subsubsection*{Disclaimer (IBM)}
IBM, the IBM logo, and ibm.com are trademarks of International Business Machines
Corp., registered in many jurisdictions worldwide. Other product and service names might be trademarks of IBM or other companies. The current list of IBM trademarks is available at https://www.ibm.com/legal/copytrade.

\bibliographystyle{unsrt}
\bibliography{main}
\end{document}